\documentstyle[preprint,aps,epsfig]{revtex}
 
\begin{document}
 
\def\real{I\negthinspace R}
\def\zed{Z\hskip -3mm Z }
\def\half{\textstyle{1\over2}}
\def\quarter{\textstyle{1\over4}}
\newcommand{\be}{\begin{equation}}
\newcommand{\ee}{\end{equation}}
\preprint{DTP/99/29, gr-qc/9904058}
\draft
\tighten
\renewcommand{\topfraction}{0.8}
 
\title{TEXTURES IN DILATON GRAVITY}
\author{Owen Dando}
\address{Centre for Particle Theory, Department of Mathematical Sciences\\
South Road, Durham, DH1 3LE}
\date{\today}
\maketitle
\begin{abstract}

We examine the field equations of a self-gravitating texture in low-energy superstring gravity, allowing for an arbitrary coupling of the texture field to the dilaton. Both massive and massless dilatons are considered. For the massless dilaton non-singular spacetimes exist, but only for certain values of the coupling, dependent on the gravitational strength of the texture; moreover, this non-singular behaviour exists only in a certain frame. For the massive dilaton, the texture induces a long-range dilaton cloud, but we expect the gravitational behaviour of the defect to be similar to that found in Einstein theory. We compare these results with those found for other global topological defects.

\end{abstract}

 
 

\section{Introduction}

Phase transitions in the early universe can give rise to a variety of topological defects which interest cosmologists as possible sources for the primordial density perturbations which seeded galaxy formation through gravitational instability \cite{BSV}. Defects are discontinuities in the vacuum and can be classified according to the topology of the vacuum manifold of the field theory model being used. Thus disconnected manifolds give domain walls, non-simply connected manifolds, strings, and vacuum manifolds with non-trivial $\Pi_2$ and $\Pi_3$ homotopy groups give monopoles and textures respectively. Defects are either local if the symmetry broken is gauged or global if it is not. The gravitational effects of both local strings \cite{LSTR} and monopoles \cite{LMON} and global domain walls \cite{GDW}, strings \cite{GSTR} and monopoles \cite{GM} have been studied extensively.

In this paper we will be concerned with texture defects, which were first proposed as possible seeds for large-scale structure formation by Turok \cite{TUROK1}. In the case of a local texture the gauge field can everywhere absorb the gradient of the texture scalar field and the defect becomes merely another vacuum configuration. Hence it is only textures arising from a broken global symmetry that will interest us. The scenario for texture formation and evolution can be put simply as follows. A non-Abelian global symmetry is broken as the early universe cools through a temperature of order the GUT-scale. This leads to a random distribution of knots, regions where the texture field winds around the vacuum manifold non-trivially. As the knots come within the cosmological horizon, the unstable texture field collapses to a very small size and unwinds, emitting a shell of Goldstone boson radiation. During their evolution, textures interact gravitationally with the cosmic fluid, inducing geometrical perturbations.

The gravitational effects of such defects were first considered by Turok and Spergel \cite{TUROK2} and N\"{o}tzold \cite{NOTZ} in the weak-field approximation. The strong gravity of textures was then investigated by Durrer \textit{et al} \cite{DUR} and Barriola and Vachaspati \cite{BV}. Guided by the fact that an exact self-similar solution to the texture field equation exists in flat space, Durrer \textit{et al} found self-similar solutions to the coupled Einstein field equations describing the metric and scalar field of a texture in curved space. The equations derived by Barriola and Vachaspati were related to those of Durrer \textit{et al} by a complicated co-ordinate transformation. Subsequently, the microwave background anisotropies induced in the texture model of structure formation have been the subject of much study (for example, see \cite{MICRO}). Over the next few years, as high resolution maps of the cosmic microwave sky become available, it should be possible to test this and other models of structure formation to high precision.

Of course, most work on defects has been performed within the context of general relativity. However, at sufficiently large energy scales it seems likely that gravity is not described solely by the Einstein action, but becomes modified, for example, by superstring terms which are scalar-tensor in nature \cite{LESG}. This low-energy superstring gravity is reminiscent of the scalar-tensor theories of Jordan, Brans and Dicke \cite{JBD}, and is in fact equivalent to Brans-Dicke theory for particular parameter values. The implications of scalar-tensor gravity for topological defects, be it Brans-Dicke or low-energy superstring theory, have been explored for local strings \cite{LSTRD}, global strings \cite{GSTRD,GSTRDUS} and global monopoles \cite{GMD,GMDUS}.

Here we will consider the implications of low-energy string gravity for global textures. We take a general form for the interaction of matter with the superstring dilaton - that is, we consider an arbitrary coupling $e^{2a\phi}{\mathcal{L}}$ between the texture lagrangian and the dilaton $\phi$. We will be concerned in particular with the questions of whether a non-singular spacetime exists for the texture, and of how the metric properties of the defect are affected by the dilaton mass. The layout of the paper is as follows. After briefly reviewing the work of Durrer \textit{et al} on textures in Einstein gravity, we present an analysis for the texture in superstring gravity, for both massless and massive dilatons. We then compare our results with those found for other global defects and present some conclusions.

\section{Textures in Einstein gravity}

In this section we will review the metric properties of textures in Einstein gravity (further details can be found in \cite{DUR}). A simple model giving rise to texture solutions is described by the Lagrangian
\be
{\mathcal{L}}(\hat{\psi}^{i})=\frac{1}{2} \nabla_{\mu} \hat{\psi}^{i} \nabla^{\mu} \hat{\psi}^{i} - \frac{\lambda}{4} (\hat{\psi}^{i} \hat{\psi}^{i} - \eta^{2})^{2} \label{FullLagr}
\ee
where $\hat{\psi^{i}}$ is a set of real scalar fields, ${i=1 \ldots 4}$. We will consider the behaviour of a single texture knot and put $\hat{\psi}=\eta \psi$. Note also that since we have not fixed $G=1$, we are free to set $\lambda \eta^2=1$ without loss of generality, which corresponds to using string rather than Planck units (now G is some small number of ${\mathcal{O}}(10^{-6})$). Then
\be
{\mathcal{L}}(\psi^{i})=\eta^2 \left ( \frac{1}{2} \nabla_{\mu} \psi^{i} \nabla^{\mu} \psi^{i} - \frac{1}{4} (\psi^{i} \psi^{i} - 1)^{2} \right ) \label{ELagr}
\ee
This model has a global $O(4)$ symmetry and the vacuum manifold is the three sphere $S^3$. The symmetry is spontaneously broken to $O(3)$ when the field $\psi^{i}$ acquires a vacuum expectation value. For grand-unified theories, $\eta$, the magnitude of this vacuum expectation value, is typically of the order $10^{16}$ GeV. Note that unlike other topological defects, the texture can remain in the vacuum manifold throughout space. In fact, the field will only leave the manifold when it has collapsed to a small enough size that its gradient energy is great enough for the texture to unwind. Hence, in order to make calculations more tractable, it is a good approximation to replace the potential term in (\ref{ELagr}) by the constraint
\be
\psi^{i} \psi^{i} = 1
\ee
The dynamics of the texture field are then determined by the Lagrangian
\be
{\mathcal{L}_{\beta}}(\psi^{i})=\eta^2 \left ( \frac{1}{2} \nabla_{\mu} \psi^{i} \nabla^{\mu} \psi^{i} - \beta (\psi^{i} \psi^{i} - 1) \right ) 
\ee
where $\beta$ is a Lagrange multiplier, giving the equation of motion
\be
\nabla_{\mu} \nabla^{\mu} \psi^{i} + \left ( \nabla_{\mu} \psi^{j} \nabla^{\mu} \psi^{j} \right ) \psi^{i} = 0
\ee
Field configurations which have a well-defined limit at spatial infinity fall into classes of the homotopy group $\Pi_3 (S^3)= \zed$, the integers of this group corresponding to winding number.

We will look for spherically-symmetric configurations describing the texture and thus make the hedgehog ansatz
\be
\psi^{i}= \left ( \cos \chi \left (r,t \right ),\sin {\chi \left (r,t \right )} \sin \theta \cos \xi ,\sin \chi \left (r,t \right ) \sin \theta \sin \xi ,\sin \chi \left (r,t \right ) \cos \theta \right ) \label{hedgehog}
\ee
There are a number of equivalent gauges in which we could discuss solutions. We will choose to write the metric
\be
ds^2=e^{2\gamma (r,t)} \left ( dt^2-dr^2 \right ) -r^2 \omega (r,t)^2 \left (d\theta^2+\sin^2 \theta d \xi^2 \right ) \label{metric}
\ee
In terms of the new variable $\chi$ the Lagrangian becomes
\be
{\mathcal{L}}=\eta^2 \left ( \half e^{-2\gamma} \left ( \dot{\chi}^2 - {\chi ' }^2 \right ) - \displaystyle{\frac{\sin ^2 \chi}{r^2 \omega^2}} \right )
\ee
where the dot and dash denote differentiation with respect to $t$ and $r$. The equation of motion for $\chi$ is 
\be
r^2 \left ( \omega^2 \dot{\chi} \right )\dot{ } - \left (r^2 \omega^2 \chi ' \right )' + e^{2\gamma} \sin 2\chi = 0
\ee
In flat space this reduces to
\be
\ddot{\chi}- \chi '' - \frac{2}{r}\chi ' + \frac{\sin 2 \chi}{r^2} = 0
\ee
which has the exact self-similar solution for $t<0$
\be
\chi=2 \arctan (-r/t) \label{flat}
\ee
describing texture collapse. Note that as $-r/t \rightarrow \infty$, $\chi \rightarrow \pi$ and $\psi$ has the constant limit $\psi=(-1,\bf{0})$ corresponding to winding number one.

The texture couples to the metric via its energy-momentum tensor
\be
G_{\mu\nu}=8 \pi G T_{\mu\nu}
\ee
Rescaling the energy-momentum tensor, $\hat{T}_{\mu\nu}=T_{\mu\nu}/\eta^2$, we have
\begin{eqnarray}
\hat{T}_t^t & = & \half e^{-2\gamma} \left ( \dot{\chi}^2 + {\chi '}^2 \right ) + \displaystyle{\frac{\sin^2 \chi}{r^2 \omega^2}} \nonumber \\
\hat{T}_t^r & = & -e^{-2\gamma} \dot{\chi} \chi ' \nonumber \\
\hat{T}_r^r & = & - \half e^{-2\gamma} \left ( \dot{\chi}^2 + {\chi '}^2 \right ) + \displaystyle{\frac{\sin^2 \chi}{r^2 \omega^2}} \nonumber \\
\hat{T}_{\theta}^{\theta} & = & - \half e^{-2\gamma} \left ( \dot{\chi}^2 - {\chi '}^2 \right ) 
\end{eqnarray}
After a little manipulation, Einstein's equations reduce to
\begin{eqnarray}
\omega \left ( \ddot{\omega} - \omega '' \right ) - \frac{4\omega \omega '}{r} + \frac{1}{r^2} \left ( e^{2\gamma}- \omega^2 \right ) + \dot{\omega}^2 - {\omega '}^2 & = & \frac{\epsilon e^{2\gamma} \sin^2 \chi}{r^2} \label{EinstOne} \\
{( r \omega )' } \, \dot{ } - \gamma' {(r \omega)}\dot{ }- \dot{\gamma} (r \omega)' & = & - \half \epsilon (r \omega) \dot{\chi} \chi ' \label{EinstTwo} \\
\frac{1}{r \omega} \left ( (r \omega)'' - (r \omega)\dot{ } \, \dot{ } \right ) + \gamma'' - \ddot{\gamma} & = & \half \epsilon \left ( \dot{\chi}^2 - {\chi '}^2 \right ) \label{EinstThree}
\end{eqnarray}
where $\epsilon=8 \pi G \eta^2$ measures the gravitational strength of the texture. In a GUT scenario $\epsilon \approx 10^{-5}$.

Guided by the fact that a self-similar solution to the texture field equation exists in flat space-time, we make the self-similar ansatz
\be
\chi (r,t)=\chi (x), \ \gamma(r,t)=\gamma(x), \ \omega (r,t)=\omega (x) 
\ee
where $x=-r/t$. If a dash now denotes differentiation with respect to $x$, the gravitational field equations become
\begin{eqnarray}
{(x\omega)'}^2 - (x^2\omega ')^2 + (x\omega)(x\omega)''(1-x^2) & = & e^{2\gamma} (1-\epsilon \sin^2 \chi) \label{SEinstOne} \\
(x\omega)'' -\gamma' (2x\omega' + \omega) & = & -\half \epsilon (x\omega) {\chi '}^2 \label{SEinstTwo} \\
\frac{(x \omega)''}{x \omega} + \frac{(\gamma' (1-x^2))'}{1-x^2} & = & - \half \epsilon \chi'^2 \label{SEinstThree}
\end{eqnarray}
and the equation of motion for $\chi$ is
\be
(x^2-1) (x^2\omega^2 \chi')' + e^{2\gamma} \sin 2\chi = 0
\ee
Note however that (\ref{SEinstTwo}) and (\ref{SEinstThree}) imply
\be
\frac{(\gamma'(1-x^2))'}{\gamma'(1-x^2)} + \frac{2\omega'}{\omega} + \frac{1}{x} = 0
\ee
which can be immediately integrated to give 
\be
\gamma'= \frac{b}{x(1-x^2)\omega^2}
\ee
where $b$ is some constant. If we make the assumption that $\gamma'$ should be finite at $x=0$ (i.e. when $r=0$ or as $t \rightarrow -\infty$) since we do not expect strong curvature effects here except at the instant the texture unwinds, we must have $b=0$ and so $\gamma=const$. Without loss of generality, we set $\gamma=0$. Hence we finally obtain the coupled equations relating the metric and texture fields $\omega$ and $\chi$ 
\begin{eqnarray}
(x \omega)'' + \half \epsilon (x \omega) {\chi '}^2 & = & 0 \label{OmegaEq} \\
(x^2-1) (x^2 \omega^2 \chi')' + \sin 2\chi & = & 0 \label{ChiEq}
\end{eqnarray}
together with the first-order constraint equation
\be
(x\omega)'^2 - (x^2\omega')^2 + \half \epsilon (x^2-1) (x\omega \chi')^2 = 1- \epsilon \sin^2 \chi \label{EinsteinConstraint}
\ee

We can easily obtain a linearised solution to these equations for the collapsing texture by expanding around the self-similar flat-space solution
\be
\chi(x) = 2 \arctan x
\ee
giving, to $O(\epsilon)$
\be
\omega (x) \approx 1+ \epsilon \left ( \frac{\arctan (x)}{x} - 1 \right )
\ee
with the boundary condition $\omega \rightarrow 1$ as $t \rightarrow -\infty$. To construct numerical solutions to the field equations we need the behaviour near $x=0$. We must have $\chi (0)=0$ or $\psi$ would be singular at the origin. Then requiring that $\omega (0)$ and $\omega '' (0)$ are finite we obtain $\omega' (0)=0$ from (\ref{OmegaEq}) and $\omega (0)=1$ from the constraint equation (\ref{EinsteinConstraint}). The field equations then imply the expansions
\begin{eqnarray}
\omega (x) & = & 1+ l x^2 + m x^4 + O (x^5) \nonumber \\
\chi (x) & = & \chi ' (0) x + n x^3 + O(x^5)  
\end{eqnarray}
where
\begin{eqnarray}
l & = & - \frac{1}{12} \epsilon \chi ' (0) ^2 \nonumber \\
m & = & -\frac{1}{100} \epsilon \chi ' (0) ^ 2 \left ( 3 + \left ( \frac{19 \epsilon}{24} -2 \right ) \chi ' (0) ^2 \right ) \\
n & = & \frac{1}{15} \chi ' (0) ( 3+ (\epsilon -2) \chi ' (0)^2 ) \nonumber
\end{eqnarray}
The only parameter in the expansions is $\chi'(0)$ which is determined as follows. The equation of motion for $\chi$ (\ref{ChiEq}) has a critical point at $x=1$ and $\chi ''$ remains finite here only if $\sin 2\chi = 0$. Hence for any given $\epsilon$ the non-singular solution is found by determining $\chi'(0)$ such that $\chi (1)=\pi/2$. 

\begin{figure}
\begin{center}
\epsfig{figure=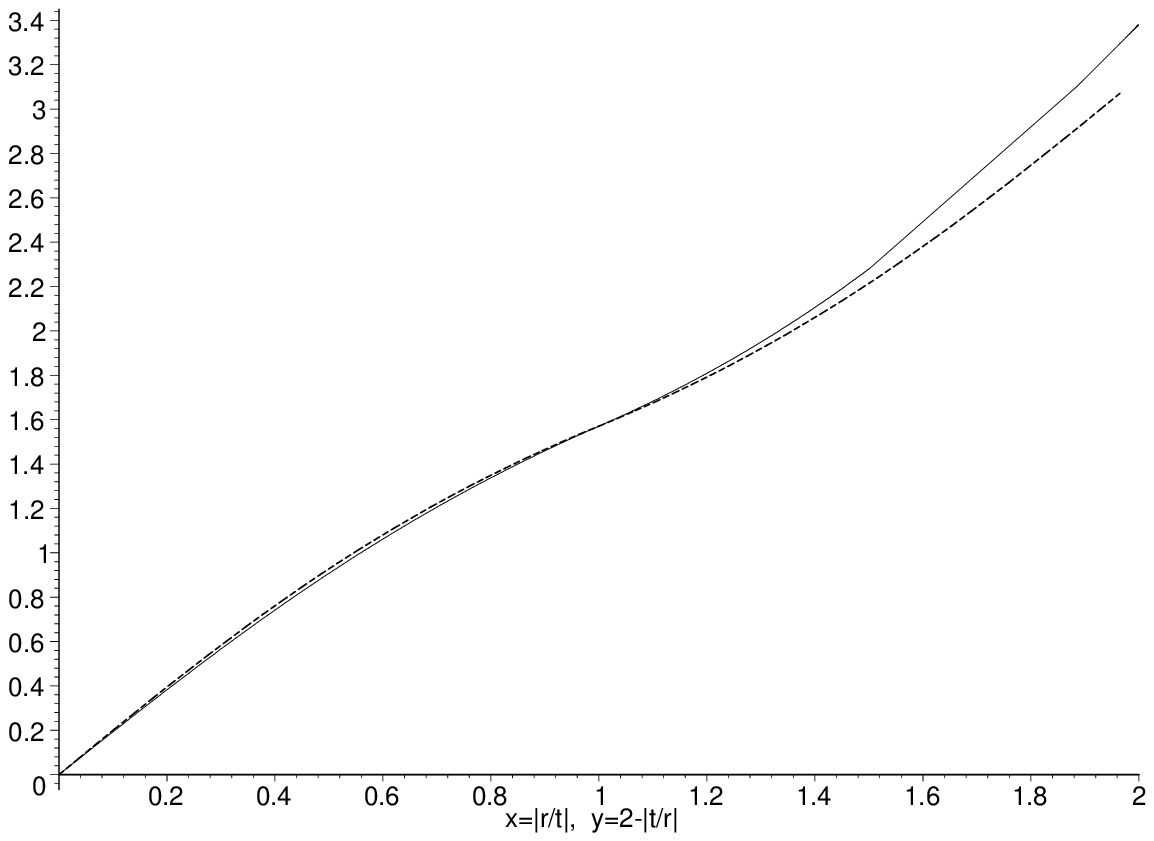,height=8cm}
\caption{The texture field $\chi$ for $\epsilon = 10^{-1}$ (solid line) together with the flat space solution $\chi=2 \arctan (-r/t)$ (dotted line). Note that the whole collapsing texture solution $0 < -r/t < \infty$ is shown. For $\epsilon >0$ the asymptotic value $\chi(\infty)$ overshoots the flat space value $\pi$. \label{fig:pic1}}
\end{center}
\end{figure}

\begin{figure}
\begin{center}
\epsfig{figure=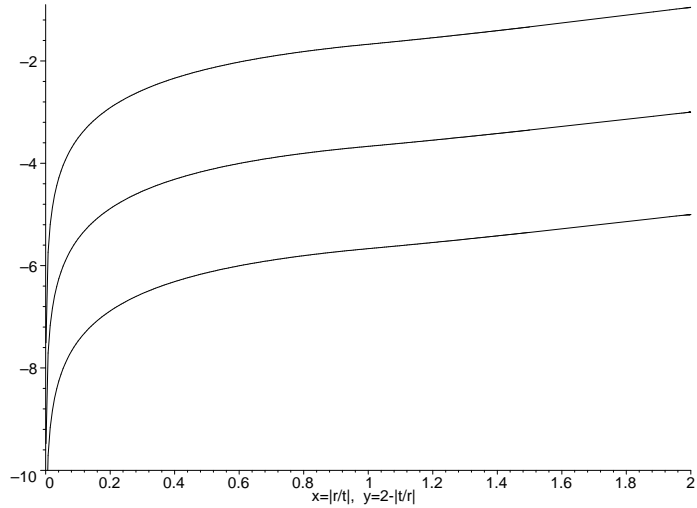,height=8cm}
\caption{The metric function $\log (1-\omega)$ for $\epsilon=10^{-1}$, $10^{-3}$ and $10^{-5}$. $\omega$ remains close to its linearised approximation for all $\epsilon$. \label{fig:pic2}}
\end{center}
\end{figure}

\begin{figure}
\begin{center}
\epsfig{figure=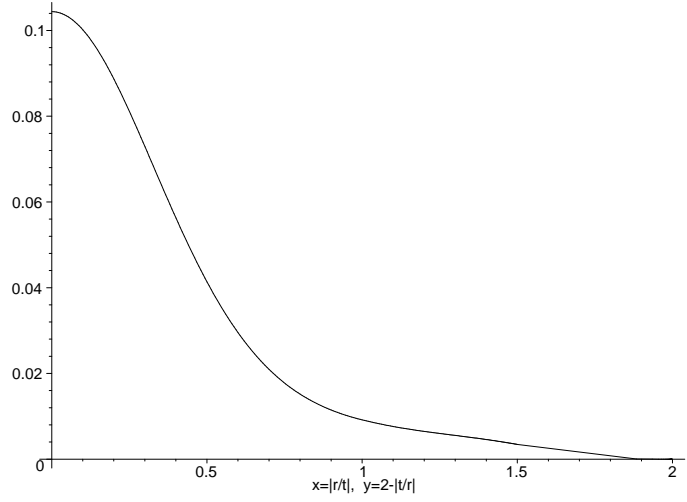,height=8cm}
\caption{The time-independent curvature invariant $\hat{R}^2_{\mu \nu \rho \sigma}=t^4 R^2_{\mu \nu \rho \sigma}$ for $\epsilon=10^{-1}$. \label{fig:pic3}}
\end{center}
\end{figure}

Figure \ref{fig:pic1} shows the amplitude of $\psi$ for the collapsing texture when $\epsilon=10^{-1}$ together with the flat-space solution $\chi (x)= 2 \arctan x$. Note that this and all subsequent figures are plotted for $0<x=-r/t<1$ and $1<y=2+t/r<2$, and hence the whole of the collapsing texture solution $0<-r/t<\infty$ is shown. Even for the unreasonably large value of $\epsilon=10^{-1}$, $\chi$ remains close to the flat-space solution. For any  $\epsilon >0$ the asymptotic value $\chi(\infty)$ always slightly overshoots the flat-space value $\pi$ and hence $\psi$ no longer has a well-defined limit at spatial infinity. The metric function $\omega$ is shown in Figure \ref{fig:pic2} for $\epsilon=10^{-1}$, $10^{-3}$ and $10^{-5}$ and is close to its linearised approximation for all values of $\epsilon$. Figure \ref{fig:pic3} shows the time-independent quantity 
\be
\hat{R}^2_{\mu \nu \rho \sigma}=t^4 R^2_{\mu \nu \rho \sigma} \propto \left ( \frac{1-{(x \omega)'}^2+ (x^2 {\omega}')^2}{x^2 \omega^2} \right ) ^2 + 2 \left ( \frac{(1-x^2)(x\omega)''}{x\omega} \right ) ^2 \label{Curve}
\ee
for $\epsilon=10^{-1}$.

The solutions describe a collapsing texture which unwinds at $t=0$. The spacetime geometry has a simple interpretation. $\omega$ changes rapidly around the light-cone $x=1$, and is almost constant, $\omega=1$ for $x \ll 1$ and $\omega \approx 1-\epsilon$ for $x \gg 1$. Hence we have a shell of rapidly changing $\omega$ moving inwards at the speed of light. Inside the shell space is nearly flat, and outside it can be approximated by a space with constant deficit solid angle $\Delta \approx \epsilon$. During the actual unwinding event $(r,|t|) \le (1/\eta,1/\eta)$, the texture leaves the vacuum manifold, and hence the sigma-model approximation is inappropriate. However, the solutions after collapse ($t>0$) can be approximated by patching together sigma-model solutions such that the winding number vanishes after collapse (see \cite{DUR}), and $\omega$ now describes the curvature effects of an expanding shell of Goldstone boson radiation.

\section{Textures in dilaton gravity}

We are interested in the behaviour of the texture metric when gravitational interactions take a form typical of low-energy string theory. In its minimal form, string gravity replaces the gravitational constant $G$ by a scalar field, the dilaton. To account for the unknown coupling of the dilaton to matter, we choose the action 
\be
S=\int d^4 x \sqrt{-\hat{g}} \left [ e^{-2\phi} \left ( -\hat{R} - 4 (\hat{\nabla} \phi)^2 - \hat{V} (\phi) \right ) + e^{2a\phi} {\mathcal{L}} \right ]
\ee
where ${\mathcal{L}}$ is as in (\ref{ELagr}) and $a$ is an arbitrary constant. The dilaton potential $\hat{V}(\phi)$ is for the moment assumed general. Note that this action is written in terms of the string metric which appears in the string sigma model. To facilitate comparison with the previous section we instead choose to write the action in terms of the `Einstein' metric
\be
g_{\mu \nu}=e^{-2\phi} \hat{g}_{\mu \nu}
\ee
in which the gravitational part of the action appears in the normal Einstein form
\be
S= \int d^4 x \sqrt{-g} \left [ -R + 2 (\nabla \phi) ^2 -V(\phi) + e^{2(a+2)\phi} {\mathcal{L}} (\psi^i,e^{2\phi} g) \right ]
\ee
where $V(\phi)=e^{2\phi} \hat{V} (\phi)$. As before, we expect the texture will only be forced to leave the vacuum manifold when it has collapsed to a very small size, and so impose $\psi^i \psi^i =1$. The dynamics of the texture field are then determined by the Lagrangian 
\be
{\mathcal{L}}_{\beta}= \half e^{2(a+1)\phi} {\nabla}_{\mu} {\psi}^i {\nabla}^{\mu} {\psi^i}  -  \beta ({\psi}^i {\psi}^i-1)
\ee
giving the equation of motion
\be
\nabla_{\mu} \nabla^{\mu} \psi^i + 2(a+1) \nabla_{\mu} \phi  \nabla^{\mu} \psi^i  + (\nabla_{\mu} \psi^j \nabla^{\mu} \psi^j) \psi^i=0
\ee
Einstein's equation becomes
\be
G_{\mu \nu}=\half e^{2(a+2)\phi} T_{\mu \nu} + S_{\mu \nu}
\ee
where the energy-momentum tensor is now
\be
T_{\mu \nu}= 2 \frac{\delta {\mathcal{L}}(\psi^i , e^{2\phi}g)}{\delta g^{\mu \nu}} - g_{\mu \nu} {\mathcal{L}}(\psi^i , e^{2\phi}g) = e^{-2\phi} \nabla_{\mu} \psi^i \nabla_{\nu} \psi^i - g_{\mu \nu} {\mathcal{L}}
\ee
and
\be
S_{\mu \nu} = 2 \nabla_{\mu} \phi \nabla_{\nu} \phi + \half g_{\mu \nu} V(\phi) - g_{\mu \nu} (\nabla \phi)^2
\ee
is the energy-momentum tensor of the dilaton, which has the equation of motion 
\be
- \Box \phi = \frac{1}{4} \frac{\partial V}{\partial \phi} - \frac{a+2}{2} e^{2(a+2)\phi} {\mathcal{L}} (\psi^i,e^{2\phi}g) + \frac{1}{4} e^{2(a+1)\phi} g^{\mu \nu} \nabla_{\mu} \psi^i \nabla_{\nu} \psi^i
\ee

As before, we will look for spherically-symmetric configurations describing the texture (\ref{hedgehog}) and choose the metric to take the form (\ref{metric}). Then the Lagrangian is
\be
{\mathcal{L}} = \eta^2 e^{-2\phi} \left ( \half e^{-2\gamma} (\dot{\chi}^2-{\chi ' }^2) -\displaystyle{\frac{\sin^2 \chi}{r^2 \omega ^2}} \right )
\ee
The rescaled, modified energy-momentum tensor, $\hat{T}_{\mu \nu}=e^{2(a+2)\phi}T_{\mu \nu}/\eta^2$, is given by
\begin{eqnarray}
\hat{T}_t^t & = & e^{2(a+1)\phi} \left (\half e^{-2\gamma} \left ( \dot{\chi}^2 + {\chi '}^2 \right ) + \displaystyle{\frac{\sin^2 \chi}{r^2 \omega^2}} \right ) \nonumber \\
\hat{T}_t^r & = & -  e^{2(a+1)\phi} e^{-2\gamma} \dot{\chi} \chi ' \\
\hat{T}_r^r & = & -  e^{2(a+1)\phi} \left ( \half e^{-2\gamma} \left ( \dot{\chi}^2 + {\chi '}^2 \right ) - \displaystyle{\frac{\sin^2 \chi}{r^2 \omega^2}} \right ) \nonumber \\
\hat{T}_{\theta}^{\theta} & = & - \half  e^{2(a+1)\phi} e^{-2\gamma} \left ( \dot{\chi}^2 - {\chi '}^2 \right )
\end{eqnarray}
and the Einstein equations may be conveniently written
\begin{eqnarray}
\omega \left ( \ddot{\omega} - \omega '' \right ) - \frac{4\omega \omega '}{r} + \frac{1}{r^2} \left ( e^{2\gamma}- \omega^2 \right ) + \dot{\omega}^2 - {\omega '}^2 & = & \frac{\epsilon e^{2(a+1)\phi} e^{2\gamma} \sin^2 \chi}{r^2} + \half e^{2\gamma} \omega^2 V(\phi) \label{DilEinstOne} \\
{( r \omega )' }^{.} - \gamma' {(r \omega)}^{.}- \dot{\gamma} (r \omega)' & = & - \half  (r \omega ) \left (\epsilon e^{2(a+1)\phi} \dot{\chi} \chi ' + 2 \dot{\phi}  \phi ' \right ) \label{DilEinstTwo} \\
\frac{1}{r \omega} \left ( (r \omega)'' - (r \omega)^{..} \right ) + \gamma'' - \ddot{\gamma} & = & \half \epsilon e^{2(a+1)\phi} \left ( \dot{\chi}^2  - {\chi '}^2 \right ) \nonumber \\
& & + \dot{\phi}^2 - {\phi '}^2 - \half e^{2\gamma} V(\phi) \label{DilEinstThree}
\end{eqnarray}
where $\epsilon=\eta^2/2$. The equation of motion for the texture field $\chi$ is 
\be
r^2 (\omega ^2 \dot{\chi})\dot{ } - (r^2 \omega^2 \chi ')' + 2(a+1)r^2 \omega^2 (\dot{\phi} \dot{\chi} - \phi ' \chi ') +e^{2\gamma} \sin 2\chi = 0
\ee
and the dilaton equation is
\be
r^2 (\omega ^2 \dot{\phi} )\dot{} - (r^2 \omega^2 \phi')' + \frac{1}{4} e^{2\gamma}r^2 \omega^2 \frac{\partial V}{\partial \phi} -\epsilon (a+1) e^{2(a+1)\phi} \left ( \half r^2 \omega^2 (\dot{\chi}^2-{\chi'}^2) -e^{2\gamma} \sin^2 \chi \right ) = 0 \label{FullDil}
\ee
We will consider massless and massive dilatons in turn.

\subsection{Massless dilatonic gravity}

For the massless dilaton $V(\phi)=0$. As in the Einstein case, it is consistent to make the self-similar ansatz
\be
\chi (r,t)=\chi (x), \ \gamma(r,t)=\gamma(x), \ \omega (r,t)=\omega (x), \ \phi (r,t)=\phi (x)
\ee
and rewrite the equations in terms of the new variable $x=-r/t$. Again, it is simple to show that on the assumption that $\gamma$ is regular at $x=0$, we must have $\gamma'=0$ where $'$ denotes differentiation with respect to $x$. Hence the coupled equations for the metric, texture and dilaton fields reduce to 
\begin{eqnarray}
(x\omega)'' + (x\omega) (\half \epsilon e^{2(a+1)\phi} {\chi'}^2 + {\phi '}^2) & = & 0  \label{MasslessDilaton1} \\
(x^2-1) (x^2 \omega^2 \chi ')' +2(a+1) x^2 \omega^2 \phi' \chi ' (x^2-1) + \sin  2\chi & = & 0 \label{MasslessDilaton2} \\
(x^2-1) (x^2 \omega^2 \phi' )' +\epsilon (a+1) e^{2(a+1)\phi} \left ( -\half x^2 \omega^2 {\chi'}^2 (x^2-1) +\sin^2 \chi \right ) & = &  0 \label{MasslessDilaton3}
\end{eqnarray}
together with the first-order constraint equation
\be
{(x \omega)'}^2 - (x^2 \omega')^2 + (x^2-1)(x\omega)^2 (\half \epsilon e^{2(a+1)\phi} {\chi'}^2 +{\phi'}^2) = 1-\epsilon e^{2(a+1)\phi} \sin^2 \chi \label{MasslessConstraint}
\ee

We can find linearised solutions to the coupled field equations by expanding around the flat-space, self-similar solution. To $O(\epsilon)$
\begin{eqnarray}
\omega (x)  & \approx & 1 + \epsilon e^{2(a+1)\phi_0} \left ( \frac{\arctan x}{x} -1 \right ) \\
\phi (x) & \approx & \phi_0 + \epsilon \left ( c_1 + \frac{c_2}{x} + (a+1)e^{2(a+1)\phi_0} \times  \right . \nonumber \\
& & \left . \left ( \ln \sqrt{ \left | \frac{1+x^2}{1-x^2} \right | } - \frac{1}{x} \left ( \arctan x + \ln \sqrt{\left | \frac{1+x}{1-x} \right | } \right ) \right ) \right )
\end{eqnarray}
where $\phi_0$, $c_1$ and $c_2$ are constants. We expect space to be flat before texture collapse $t \rightarrow -\infty$, and during collapse at $r=0$. Hence we will impose $\phi=0$ when $x=0$. This fixes the constants $\phi_0$ and $c_i$ and we obtain
\begin{eqnarray}
\omega(x) & \approx & 1 +\epsilon \left ( \frac{\arctan x}{x} - 1\right ) \label{LinDilOmega} \\
\phi(x) & \approx & \epsilon (a+1) \left ( 2+ \ln \sqrt{\left | \frac{1+x^2}{1-x^2} \right |} - \frac{1}{x} \left ( \arctan x + \ln \sqrt{\left | \frac{1+x}{1-x} \right |} \right ) \right ) \label{LinDilDil}
\end{eqnarray}

Note that if $a=-1$ then $\phi=0$ to first order. Although (\ref{MasslessDilaton1}) includes the dilaton field even if $a=-1$, in this case a trivial dilaton $\phi=0$ satisfies the boundary conditions and the equations are reduced to those seen in Einstein gravity. This is in contrast to the behaviour of other global topological defects in string gravity (see \cite{GSTRDUS,GMDUS}) where the dilaton is non-trivial for $a=-1$, and is due to the texture field being constrained to remain in the vacuum manifold. Note also that although $\phi$ is continuous in the linearised approximation, $\phi'$ diverges at $x=1$. This is an indication of qualitatively different behaviour that will make the solution of the full field equations slightly more complicated than in the Einstein case.

We attempt to find numerical solutions as before. We first need the behaviour of the fields near $x=0$. Taking $\phi(0)=0$, again we must have $\chi(0)=0$ or $\psi$ would be singular at the origin. Then requiring that $\omega(0)$, $\omega''(0)$ and $\phi'(0)$ are finite implies $\omega'(0)=0$ from (\ref{MasslessDilaton1}). Thus from the constraint equation (\ref{MasslessConstraint}), $\omega(0)=1$ and we obtain the expansions
\begin{eqnarray}
\omega (x) & = & 1 + l x^2 + m x^4 + O(x^5) \nonumber \\
\chi (x) & = & \chi' (0) x + n x^3 + O(x^5) \\
\phi (x) & = & p x^2 + q x^4 + O(x^5) \nonumber
\end{eqnarray}
where 
\begin{eqnarray}
l & = & -\frac{1}{12} \epsilon \chi'(0)^2 \nonumber \\
m & = & -\frac{1}{100} \epsilon \chi'(0)^2 \left ( 3+ \left ( \epsilon \left ( k^2 + \frac{19}{24} \right ) -2 \right ) \chi'(0)^2 \right ) \nonumber \\
n & = & \frac{1}{15} \chi'(0) \left ( 3+ \left ( \epsilon \left ( 1-\frac{3}{2} k^2 \right ) -2 \right ) \chi'(0)^2 \right ) \\
p & = & \frac{1}{4} \epsilon k \chi'(0)^2 \nonumber \\
q & = & \frac{1}{80} \epsilon k \chi'(0)^2 \left (8+ \left ( \epsilon \left ( k^2 + \frac{8}{3} \right ) -4 \right ) \chi'(0)^2 \right ) \nonumber
\end{eqnarray}
and $k=a+1$. As before, the only free parameter in the expansions is $\chi'(0)$ and the equation of motion for $\chi$ (\ref{MasslessDilaton2}) has a critical point at $x=1$. Note however that the dilaton equation (\ref{MasslessDilaton3}) also has a critical point here and we cannot simply set $\chi(1)=\pi/2$ to ensure the continuity of all derivatives at $x=1$. In fact, as can be seen from the field equations, only one of $\chi$ and $\phi$ can be differentiable at $x=1$. We choose to adjust $\chi'(0)$ so that $\chi'$ remains finite to facilitate comparisons with the Einstein solutions. Then $\phi'$ will not remain finite as $x\rightarrow 1$ and, as can be seen from (\ref{MasslessDilaton1}), neither will $\omega''$. Simple numerical integration will fail at $x=1$.

Instead, we integrate from $x=0$ to $x^{-}$ close to $x=1$. Using $\omega''(x^{-})$ and $\chi''(x^{-})$ obtained from the field equations, we can estimate the values of $\chi$ and $\omega$ and their derivatives and $\phi$ at $x=1$, remembering that $\phi'$ is not finite here. Again, we adjust $\chi'(0)$ so that $\chi(1) = \pi/2$. Rewriting the field equations in terms of $v=1/x$,
\begin{eqnarray}
\ddot{\omega} + \omega \left ( \half \epsilon e^{2(a+1)\phi} {\dot{\chi}}^2 + {\dot{\phi}}^2 \right ) & = & 0 \\
(\omega^2 \dot{\chi}) \dot{\ } + 2(a+1) \omega^2 \dot{\phi} \dot{\chi} + \frac{\sin 2\chi}{1-v^2} & = & 0 \\
{(\omega^2 \dot{\phi})} \dot{\ } + \epsilon (a+1) e^{2(a+1)\phi} \left ( -\half \omega^2 \dot{\chi}^2 + \displaystyle{\frac{\sin^2 \chi}{1-v^2}} \right ) & = & 0
\end{eqnarray}
where a dot indicates differentiation with respect to $v$, we can integrate from $v=0$ to $v^{-}=1/x^{-}$ and estimate the values of the variables at $v=1$. Then ensuring that the values obtained from both integrations for $\omega$, $\omega'$, $\chi$, $\chi'$ and $\phi$ match at $x=v=1$ we should have obtained complete solutions to the field equations such that $\chi$ and $\omega$ are continuous and once-differentiable and $\phi$ is continuous but not differentiable at $x=1$. Of course, some error is involved in the estimates we have made, but the results are reasonably robust to changes in $x^{-}$.

\begin{figure}
\begin{center}
\epsfig{figure=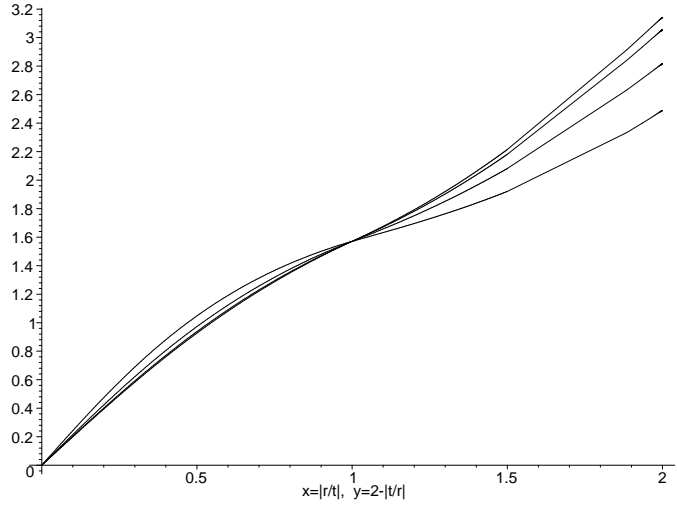,height=8cm}
\caption{The texture field $\chi$ for $\epsilon=10^{-3}$ and $|a+1|=0,5,10,15$. As $|a+1|$ increases the asymptotic value of $\chi$ decreases and we see progressively larger deviations from the Einstein theory solution $a=-1$. \label{fig:pic4}}
\end{center}
\end{figure}

\begin{figure}
\begin{center}
\epsfig{figure=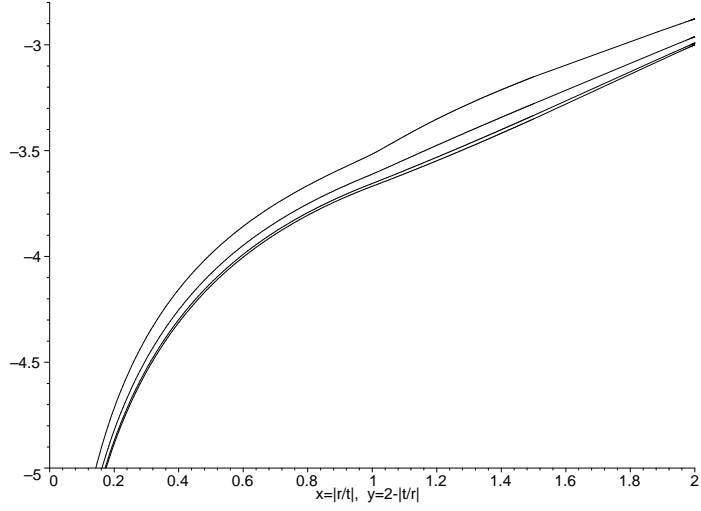,height=8cm}
\caption{The metric function $\log (1-\omega)$ for $\epsilon=10^{-3}$ and $|a+1|=0,5,10,15$. $\log (1-\omega)$ increases with $|a+1|$ and the solid deficit angle of the asymptotic spacetime is progressively greater. \label{fig:pic5}}
\end{center}
\end{figure}

\begin{figure}
\begin{center}
\epsfig{figure=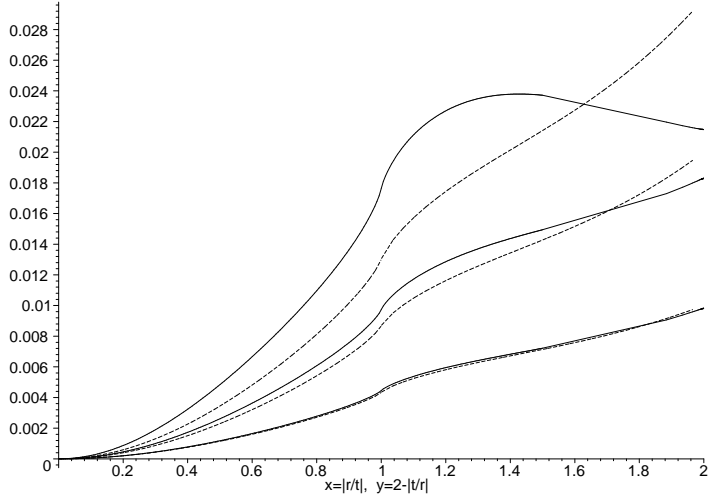,height=8cm}
\caption{The dilaton field $|\phi|$ for $\epsilon=10^{-3}$ and $|a+1|=5,10,15$ (solid lines) together with the linearised approximations (dotted lines). $|\phi|$ increases with $|a+1|$ and we see progressively larger deviations from the linearised solutions. \label{fig:pic6}}
\end{center}
\end{figure}

\begin{figure}
\begin{center}
\epsfig{figure=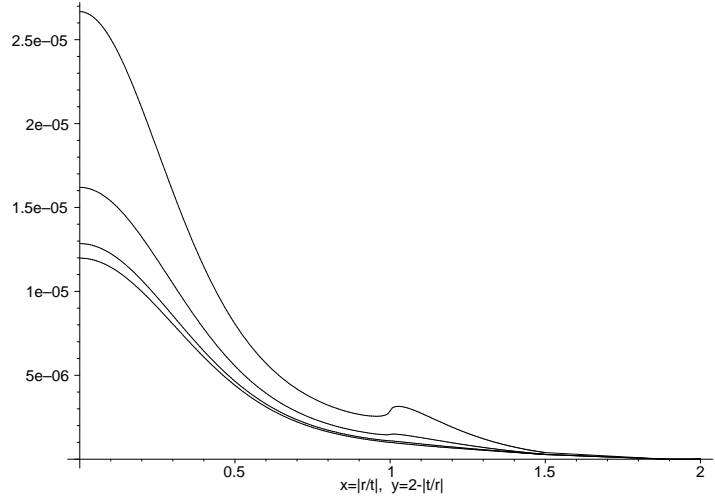,height=8cm}
\caption{The curvature invariant $\hat{R}^2_{\mu \nu \rho \sigma}=t^4 R^2_{\mu \nu \rho \sigma}$ for $\epsilon=10^{-3}$ and $|a+1|=0,5,10,15$.  As $|a+1|$ increases, $\hat{R}^2_{\mu \nu \rho \sigma}$ increases globally, and a kink develops around the hypersurface $x=1$. \label{fig:pic7}}
\end{center}
\end{figure}

Figures \ref{fig:pic4} and \ref{fig:pic5} show respectively the amplitude of $\psi$ and the metric function $\omega$ for the collapsing texture for $\epsilon=10^{-3}$ and $|a+1|=5,10$ and $15$ together with the solutions found in Einstein gravity ($a=-1$). Figure \ref{fig:pic6} shows the modulus of the dilaton field $|\phi|$ for $\epsilon=10^{-3}$ and $|a+1|=5,10$ and $15$, together with the linearised approximation (\ref{LinDilDil}). Finally, Figure \ref{fig:pic7} shows the time-independent quantity $\hat{R}^2_{\mu \nu \rho \sigma}=t^4 R^2_{\mu \nu \rho \sigma}$ for the same solutions. Under the sign change $a+1 \rightarrow -(a+1)$, the solutions for $\chi$ and $\omega$ are unaltered, but $\phi \rightarrow -\phi$. 

Note that as $|a+1|$ increases, the gradient of $\chi$ at $x=1$ progressively decreases and that $\hat{R}^2_{\mu \nu \rho \sigma}$ develops a `kink' here, an indication that curvature effects are increasing on the light-cone. In fact, beyond a certain value of $|a+1|$, our numerical procedure fails and such solutions can no longer be found. We will now show that for any $\epsilon$, there is a value of $|a+1|$ beyond which non-singular texture solutions do not exist in massless dilatonic gravity.

Consider the equation of motion for the texture field $\chi$ (\ref{MasslessDilaton2}) which we rewrite as
\be
\frac{(x^2 \omega^2 {\chi}')'}{x^2 \omega^2 \chi'} = -2(a+1){\phi}' + \frac{\sin 2\chi}{x^2 (1-x^2) \omega^2 \chi'}
\ee
For some small $\kappa$, with $1 \ge x > \kappa$, we can integrate this to give
\be
\ln(x^2 \omega^2 \chi') - \ln({\kappa}^2 \omega (\kappa)^2 \chi'(\kappa))  = -2(a+1)(\phi (x)-\phi (\kappa)) + \int_{\kappa}^{x} \frac{\sin 2\chi}{\hat{x}^2 (1-\hat{x}^2) \omega^2 \chi'} d\hat{x} 
\ee
Then taking exponentials and letting $\kappa \rightarrow 0$ gives
\be
\chi'(x) = \frac{1}{x^2 \omega^2} \exp \left ( -2(a+1)\phi + \int_{0}^{x} \frac{\sin 2\chi}{\hat{x}^2 (1-\hat{x}^2) \omega^2 \chi'} d\hat{x} \right ) \label{TextureInt}
\ee
Suppose that $\chi'$ decreases to zero at some $x_0 \le 1$. Then given that the integrand in (\ref{TextureInt}) is positive on the interval $[0,x_0]$, we must either have $\exp(2(a+1)\phi)\rightarrow \infty$ or $|\omega| \rightarrow \infty$ as $x \rightarrow x_0$. Now if $\omega \rightarrow 0$ on $[0,x_0]$, then $\exp(2(a+1)\phi) \rightarrow \infty$ for $\chi'$ to remain finite. But (\ref{MasslessDilaton1}) implies $\omega \le 1$, so we definitely must have $\exp(2(a+1)\phi) \rightarrow \infty$ as $x \rightarrow x_0$. Then consider the constraint equation (\ref{MasslessConstraint}) which we write as
\be
-\frac{1 - {(x\omega)'}^2 + (x^2 \omega')^2}{x^2\omega^2}+ \frac{(1-x^2) (x\omega)''}{x\omega} = - \frac{\epsilon e^{2(a+1)\phi} \sin^2 \chi}{x^2 \omega^2}
\ee
Since we have $0 < \chi(x_0) \le \pi/2$, the RHS tends to $- \infty$ as $x \rightarrow x_0$. So at least one of the terms on the LHS must tend to $- \infty$. In either case, the quantity $t^4 R^2_{abcd}$ (\ref{Curve}) becomes infinite as $x \rightarrow x_0$ and the spacetime is singular. 

It remains to be shown that as we increase $|a+1|$ we necessarily approach a solution for which $\chi'(1)=0$. By integrating the dilaton equation (\ref{MasslessDilaton3}) we obtain :
\be
(a+1)\phi'(x) = \frac{\epsilon (a+1)^2}{x^2 \omega (x)^2} \int_0^{x} e^{2(a+1)\phi} \left ( \half \hat{x}^2 \omega^2 {\chi'}^2 + \displaystyle{\frac{\sin^2 \chi}{1-\hat{x}^2}} \right ) d\hat{x} \label{DilatonInt}
\ee
Hence for any $x \in [0,1]$, $(a+1)\phi'(x) > 0$. Now consider the equation of motion for $\chi$ which we integrate to give :
\be
\omega(1)^2 \chi'(1) = -2(a+1) \int_0^1 x^2 \omega^2 \phi' \chi' dx + \int_0^1 \displaystyle{\frac{\sin 2\chi}{1-x^2}} dx \label{Chi1}
\ee
Clearly, the first term on the RHS is negative, whilst the second is positive. Then note that (\ref{DilatonInt}) also shows that 
\be
(a+1) x^2 \omega^2 \phi' > \half \epsilon (a+1)^2 \displaystyle{\int_0^x} e^{2(a+1)\phi} \hat{x}^2 \omega^2 {\chi'}^2 d\hat{x}
\ee
since the other term in the integrand in (\ref{DilatonInt}) is always positive on $[0,1]$. But (\ref{TextureInt}) gives
\be
x^2 \omega^2 \chi' = \exp \left ( -2 (a+1) \phi + \int_0^x \frac{\sin 2\chi}{\hat{x} (1-\hat{x}^2) \omega^2 \chi'} d\hat{x} \right ) > e^{-2(a+1)\phi}
\ee
implying
\begin{eqnarray}
(a+1) x^2 \omega^2 \phi' & > & \half \epsilon (a+1)^2 \displaystyle{\int_0^x} e^{2(a+1)\phi} \hat{x}^2 \omega^2 {\chi'}^2 d\hat{x} \nonumber \\
& > & \half \epsilon (a+1)^2 \displaystyle{\int_0^{x}} \chi' d\hat{x} = \half \epsilon (a+1)^2 \chi (x)
\end{eqnarray}
Hence
\be
2(a+1) \int_0^1 x^2 \omega^2 \phi' \chi' dx > \epsilon (a+1)^2 \int_0^1 \chi \chi' dx = \half \epsilon (a+1)^2 [\chi^2]_0^1
\ee
and since we require $\chi(1)=\pi/2$, we have
\be
2(a+1) \int_0^1 x^2 \omega^2 \phi' \chi' dx > \frac{\epsilon (a+1)^2 \pi^2}{8}
\ee

Now consider the other term on the RHS of (\ref{Chi1})
\be
\int_0^1 \frac{\sin 2\chi}{1-x^2} dx = \half \displaystyle{\int_0^1} \displaystyle{\frac{\sin 2\chi}{1+x}} dx + \half \displaystyle{\int_0^1} \displaystyle{\frac{\sin 2\chi}{1-x}} dx
\ee
For this to grow without bound as $|a+1|$ increases we would certainly need, for example
\be
\frac{\sin 2\chi}{1-x} > \frac{1}{\sqrt{1-x}}
\ee
in a neighbourhood of $x=1$, since the integral $\int_0^1 (1-x)^{-1/2}=2$ shows no such extreme behaviour. But then $\sin 2\chi > \sqrt{1-x}$ in this region, requiring 
\be
|\chi' \cos 2\chi| > \frac{1}{4\sqrt{1-x}}
\ee
As $\cos 2\chi \approx -1$, this is clearly not the case since we are looking for solutions for which $\chi'(1)$ remains finite. Hence for any value of $\epsilon$, there must be a critical value of $c=|a+1|$ above which
\be
2(a+1) \int_0^1 x^2 \omega^2 \phi' \chi' dx > \frac{\epsilon (a+1)^2 \pi^2}{8} > \int_0^1 \frac{\sin 2\chi}{1-x^2} dx
\ee
For $|a+1|>c$, we must have $\chi'(x_0)=0$ for some $x_0 \le 1$ and the texture spacetime is singular.

We can obtain a rough estimate of the critical value of $|a+1|$ by using the flat-space solution for the texture, $\chi_0=2 \arctan x$, for which
\be
\displaystyle{\int_0^1} \displaystyle{\frac{\sin 2\chi_0}{1-x^2}} dx = 1 
\ee
Then
\be
|a+1| \approx \frac{2}{\pi} \sqrt{\frac{2}{\epsilon}}
\ee
In fact the actual critical value is considerably lower than this. As we approach this value, numerical investigation shows
\be
\displaystyle{\int_0^1} \displaystyle{\frac{\sin 2\chi}{1-x^2}} dx \approx 1/2 \Rightarrow |a+1| \approx \frac{2}{\pi} \sqrt{\frac{1}{\epsilon}}
\ee
agreeing well with the values we have found.

For $\epsilon=10^{-3}$, the critical value of $|a+1|$ is 
\be
|a+1| \approx \frac{2}{\pi}\sqrt{\frac{1}{10^{-3}}} \approx 20
\ee
Figures \ref{fig:pic4} - \ref{fig:pic7} indicate that as we increase $|a+1|$ towards the critical value we see progressively larger deviations from the Einstein theory solutions. The asymptotic values of both $\chi$ and $\omega$ are reduced from their values in Einstein theory. The geometrical interpretation of solutions is as before; however, as $|a+1|$ increases, $\omega$ changes more rapidly near the light cone, and the solid deficit angle of the space outside this region is progressively greater. Note also from Figure \ref{fig:pic6} that as $|a+1|$ increases, the deviations of $\phi$ from the linearised solution are large. It is unclear how far to trust the solutions obtained for $|a+1|$ near the critical value, as errors introduced by our numerical procedure may become significant.

For the non-singular solutions we have found with $|a+1|$ less
than the critical value, $R^2_{\mu \nu \rho \sigma}$ remains finite
even though $\phi'$ and $\omega''$ diverge as $x \rightarrow 1$. Thus
the null hypersurface $x=1$ is not a scalar polynomial curvature
singularity; that is, any polynomial constructed from the scalars $R^2_{\mu \nu \rho \sigma}$, $R^2_{\mu \nu}$ and $R$ remains finite here. However, we should note that the possibility remains that this hypersurface is a non-scalar polynomial curvature singularity. 

Indeed, it seems that the `non-singular' behaviour observed
is confined to the Einstein frame. If we make the inverse transformation
$\hat{g}_{\mu \nu}=e^{2\phi} g_{\mu \nu}$ back to the string frame, the
properties of the conformal factor are included in the curvature. Even
though the metric function $\hat{\omega}=e^{\phi}\omega$ remains continuous, its
derivative diverges. Examining $R^2_{\mu \nu \rho \sigma}$ in the string frame, we see that the hypersurface $x=1$ has become a scalar polynomial curvature singularity. By examining the geodesic equations in a linearised approximation in the string frame, one can show that null and timelike geodesics reach the null hypersurface $x=1$ at finite affine parameter. We are thus able to conclude that there are no non-singular texture solutions in the string frame unless $a=-1$, at least if we impose the self-similar ansatz.

Finally, it is instructive to note that we can obtain the linearised solution (\ref{LinDilDil}) directly from (\ref{FullDil}) by a method other than direct integration of the self-similar field equations. Expanding in powers of $\epsilon$ about the flat space solution $\chi=\arctan(-r/t)$ we obtain the following equation for $\hat{\phi_1}=r \phi_1$
\be
\ddot{\hat{\phi_1}}- {\hat{\phi_1}}''= 2(a+1)r \frac{r^2-3t^2}{(r^2+t^2)^2}
\ee
This is simply the inhomogenous wave equation. The Green's function appropriate for the boundary conditions we require, that is $\hat{\phi_1} \rightarrow 0$ as $t \rightarrow -\infty$ and at $r=0$, is
\be
G(r,t;r',t')=- \half H[(r-r')^2 - (t-t')^2]
\ee 
where $H$ is the Heaviside step-function. Hence we can write down the solution for $\phi_1$ as
\begin{eqnarray}
\phi_1 & = & - \frac{2(a+1)}{r} \int_0^{\infty} dr' \int_{-\infty}^{\infty} dt' \ G(r,t;r',t') r' \frac{r'^2-3t'^2}{(r'^2+t'^2)^2} \nonumber \\
& = & - \frac{a+1}{r} \int_0^r dr' \int_{t-(r-r')}^{t+(r-r')} dt' \ r' \frac{r'^2-3t'^2}{(r'^2+t'^2)^2} 
\end{eqnarray}
This expression, valid for both $r<|t|$ and $r>|t|$, indeed gives (\ref{LinDilDil}) after some calculation. This will enable us to write down a solution in the case of the massive dilaton, to which we now turn.

\subsection{Massive dilatonic gravity}

For the massive dilaton, we take $V(\phi)=2m^2 \phi^2$ where the mass $m$ is measured in units of the Higgs mass. We do not expect this to be the exact form of the potential, but for small perturbations of the dilaton away from its vacuum value, we might expect a quadratic form to be a good approximation. We take $10^{-11} \le m \le 1$, representing a range for the unknown dilaton mass of 1 TeV - $10^{15}$ GeV. The gravitational field equations are
\begin{eqnarray}
\omega \left ( \ddot{\omega} - \omega '' \right ) - \frac{4\omega \omega '}{r} + \frac{1}{r^2}
\left ( e^{2\gamma}- \omega^2 \right ) & + & \dot{\omega}^2 - {\omega '}^2  = 
\frac{\epsilon e^{2(a+1)\phi} e^{2\gamma} \sin^2 \chi}{r^2} + m^2 e^{2\gamma} \omega^2 \phi^2
\label{MassDilEinstOne} \\
{( r \omega )' }^{.} - \gamma' {(r \omega)}^{.}- \dot{\gamma} (r \omega)' & = & - \half  (r \omega)
\left (\epsilon e^{2(a+1)\phi} \dot{\chi} \chi ' + 2 \dot{\phi}  \phi ' \right )
\label{MassDilEinstTwo} \\
\frac{1}{r \omega} \left ( (r \omega)'' - (r \omega)^{..} \right ) + \gamma'' - \ddot{\gamma} & = &
\half \epsilon e^{2(a+1)\phi} \left ( \dot{\chi}^2 - {\chi '}^2 \right ) + \dot{\phi}^2 -
{\phi '}^2 - m^2 e^{2\gamma} \phi^2 \label{MassDilEinstThree}
\end{eqnarray}
and the dilaton equation is
\be
r^2 (\omega ^2 \dot{\phi} )^{.} - (r^2 \omega^2 \phi')' + m^2 e^{2\gamma} r^2
\omega^2 \phi -\epsilon (a+1) e^{2(a+1)\phi} \left ( \half r^2 \omega^2
(\dot{\chi}^2-{\chi'}^2) -e^{2\gamma} \sin^2 \chi \right ) = 0
\ee
The presence of the $m^2 e^{2\gamma} r^2 \omega^2 \phi$ term means we can no longer reduce the equations to a system of ordinary differential equations by writing them in terms of the self-similar variable $x=-r/t$. Instead, we expand the equations about flat space as they are. Then to $O(\epsilon)$, the dilaton equation gives 
\be
\ddot{\hat{\phi}_1} - {\hat{\phi}_1}'' + m^2 \hat{\phi}_1 = 2 (a+1) r \displaystyle{\frac{r^2-3t^2}{(r^2+t^2)^2}} \label{Klein}
\ee
where $\phi=\epsilon \phi_1+\ldots$ and $\hat{\phi}_1=r\phi_1$. (\ref{Klein}) is the inhomogenous Klein-Gordon equation. From the analysis above for the massless dilaton, we know the prescription that will give the appropriate boundary conditions as $t \rightarrow -\infty$ and at $r=0$. That is, the correct Green's function is
\be
G(r,t;r',t') = - \half H[(r-r')^2-(t-t')^2] J_0 [m\sqrt{(r-r')^2-(t-t')^2}]
\ee
where $J_0$ is the Bessel function of order zero. Hence the solution for $\phi_1$ is
\be
\phi_1 = -\frac{a+1}{r} \int_0^r dr' \int_{t-(r-r')}^{t+(r-r')} dt' \ J_0 [m\sqrt{(r-r')^2-(t-t')^2}] \ r' \frac{r'^2-3t'^2}{(r'^2+t'^2)^2} \label{MassiveSol}
\ee
Note that for small argument $y$, $J_0 (y) \approx 1 - y^2/4$. Hence if $mr \ll 1$, (\ref{MassiveSol}) gives
\be
\phi_1 \approx - \frac{a+1}{r} \int_0^r dr' \int_{t-(r-r')}^{t+(r-r')} dt' \ \left ( 1-\frac{m^2}{4} \left ( (r-r')^2 -(t-t')^2 \right ) \right ) r' \frac{r'^2-3t'^2}{(r'^2+t'^2)^2}
\ee
If we further impose $r \ll t$, then to next to leading order we obtain
\begin{eqnarray}
\phi_1 & \approx & \frac{3(a+1)}{r} \int_0^r dr' \int_{t-(r-r')}^{t+(r-r')} dt' \ r' \left ( \frac{1}{t'^2} + \frac{m^2}{4} \left ( \frac{t}{t'}-1 \right )^2 \right ) \nonumber \\
& = & 3 (a+1) \left ( \left ( 2- \ln (1-x^2) - \frac{1}{x} \ln \left ( \frac{1+x}{1-x} \right ) \right ) \right . \nonumber \\
& & \left . + \frac{m^2t^2}{4} \left ( 4+ \frac{x^2}{3} -3 \ln (1-x^2) -\frac{1}{x} (2+x^2) \ln \left ( \frac{1+x}{1-x} \right ) \right ) \right ) \label{SmallRSol}
\end{eqnarray}
where $x=-r/t$. For the values of $r$ and $t$ under consideration, (\ref{SmallRSol}) remains very close to the massless solution (\ref{LinDilDil}). For $r \gg t$, where we expect $\ddot{\phi_1}, \ \phi_1'' \approx 0$, we obtain from (\ref{Klein})
\be
\phi_1 \approx \frac{2(a+1)}{m^2 r^2}
\ee
Thus the texture supports a diffuse dilaton cloud. This power law fall-off of a massive scalar field seems counterintuitive until one remembers that the dilaton is in fact part of the gravitational sector of the theory, and therefore couples to the energy-momentum of the texture. The slow fall-off of this energy-momentum is what supports the diffuse dilaton cloud. 

Finally, note that in the case of other global topological defects \cite{GSTRDUS,GMDUS}, it was found that whilst spacetimes were generically singular for the massless dilaton, for the massive dilaton they were similar to those found in Einstein gravity. Here, the presence of the dilaton seems less destructive, presumably because we have constrained the texture field to remain in the vacuum manifold, and we have shown that non-singular solutions do exist for the massless dilaton, at least for some parameter values. Hence, it might be expected that although the texture and metric fields for the massive dilaton will show some complicated non-self-similar behaviour, they should be well approximated by the Einstein theory solutions.

\section{Discussion}

We have studied the metric and dilaton fields of a global texture in low-energy string gravity for an arbitrary coupling of the texture Lagrangian to the dilaton : $e^{2a\phi}{\mathcal{L}}$. For the massless dilaton, we were able to reduce the partial differential equations describing the metric, texture and dilaton fields to ordinary differential equations in the self-similar variable $x=-r/t$. We found that in contrast to the behaviour of other global topological defects in dilaton gravity, non-singular spacetimes do exist, at least for certain parameter values. However, we have shown that for any value of $\epsilon$, which measures the gravitational strength of the texture, there exists a critical value of $|a+1|$ beyond which non-singular solutions do not exist. This critical value is approximately given by
\be
|a+1| \approx \frac{2}{\pi} \sqrt{\frac{1}{\epsilon}}
\ee
For values of $|a+1|$ for which non-singular solutions do exist, the behaviour of the metric
and texture fields is broadly similar to Einstein gravity, but the far-field spacetime shows
an increased solid deficit angle. We have noted that this `non-singular'
behaviour seems to be confined to the Einstein frame. In the string frame, the null hypersurface $x=1$ is a scalar polynomial curvature singularity and null and timelike geodesics reach this hypersurface at finite affine parameter. Thus the texture spacetime is singular in the string frame if $a\ne -1$.

For $a=-1$, the boundary conditions are satisfied by a trivial dilaton
$\phi=0$, and the field equations reduce to those found in Einstein
gravity. This is in contrast to other global topological defects \cite{GSTRDUS,GMDUS} where $\phi$ remains non-trivial when $a=-1$, and is due to
the fact that in the sigma-model approximation we have constrained the texture field to remain in the
vacuum manifold.

For the massive dilaton, the partial differential equations governing the fields cannot be simplified as they were in the massless case. However, using the appropriate Green's function, we were able to write down an integral for the dilaton in the linearised approximation that satisfies the appropriate boundary conditions. For $r \ll t$ and $mr \ll 1$ the dilaton was well approximated by the massless solution (\ref{LinDilDil}). For $r \gg t$, we found the asymptotic behaviour $\phi \approx 2 \epsilon (a+1) / m^2 r^2$. By contrasting the results obtained for the texture in massless dilatonic gravity with those for other global topological defects, we are led to suspect that the presence of a massive dilaton will do little to alter the behaviour of the texture from that found in Einstein gravity. Astrophysical bounds \cite{MICRO} on global textures obtained from their metric field will hence be unchanged by the dilaton; the conclusion that global texture models of structure formation are not favored by current observations of CMB anisotropies (see for example \cite{PENSEL}) still holds. Moreover, since the dilaton fall-off is a power law in $mr$, we expect that the Damour-Vilenkin bound \cite{DAMVIL} will hold, and global textures will be inconsistent with a low (TeV) mass dilaton.

In conclusion, we have found that in low-energy string gravity, textures show qualitatively different behaviour than in Einstein gravity, but perhaps to a lesser extent than was found for other global topological defects. 

\section*{Acknowledgements}

O.D. is supported by a PPARC studentship. I would like to thank Ruth Gregory for many helpful discussions; I would also like to thank the referee for their insightful comments.

\begin{thebibliography}{99}
 
\def\apj#1 #2 #3.{{\it Ap.\ J.\ \bf#1} #2 (#3).}
\def\cmp#1 #2 #3.{{\it Commun.\ Math.\ Phys.\ \bf#1} #2 (#3).}
\def\cqg#1 #2 #3.{{\it Class.\ Quantum Grav.\ \bf#1} #2 (#3).}
\def\grg#1 #2 #3.{{\it Gen.\ Rel.\ Grav.\ \bf#1} #2 (#3).}
\def\npb#1 #2 #3.{{\it Nucl.\ Phys.\ \rm B\bf#1} #2 (#3).}
\def\pla#1 #2 #3.{{\it Phys.\ Lett.\ \bf#1\/}A #2 (#3).}
\def\plb#1 #2 #3.{{\it Phys.\ Lett.\ \bf#1\/}B #2 (#3).}
\def\pr#1 #2 #3.{{\it Phys.\ Rev.\ \bf#1} #2 (#3).}
\def\prd#1 #2 #3.{{\it Phys.\ Rev.\ \rm D\bf#1} #2 (#3).}
\def\prl#1 #2 #3.{{\it Phys.\ Rev.\ Lett.\ \bf#1} #2 (#3).}
\def\zphys#1 #2 #3.{{\it Zeit.\ Phys.\ \bf#1} #2 (#3).}

\bibitem{BSV} A.Vilenkin and E.P.S.Shellard, {\it Cosmic strings and other Topological Defects} (Cambridge Univ. Press, Cambridge, 1994).\hfill\break
R.H.Brandenberger, {\it Modern Cosmology and Structure Formation}. [astro-ph/9411049]

\bibitem{LSTR} A.Vilenkin, \prd 23 852 1981. \hfill\break
J.R.Gott III, \apj 288 422 1985. \hfill\break
W.Hiscock, \prd 31 3288 1985. \hfill\break
B.Linet, \grg 17 1109 1985. \hfill\break
D.Garfinkle, \prd 32 1323 1985. \hfill\break
R.Gregory, \prl 59 740 1987.

\bibitem{LMON} M.Ortiz, \prd 45 2586 1992. \hfill\break
K.Lee, V.P.Nair, and E.J.Weinberg, \prd 45 2751 1992.

\bibitem{GDW} A.Vilenkin, \plb 133 177 1983. \hfill\break
J.Ipser and P.Sikivie, \prd 30 712 1984.

\bibitem{GSTR} A.G.Cohen and D.B.Kaplan, \plb 215 67 1988. \hfill\break
D.Harari and P.Sikivie, \prd 37 3438 1988. \hfill\break
R.Gregory, \plb 215 663 1988. \hfill\break
G.Gibbons, M.Ortiz and F.Ruiz, \prd 39 1546 1989. \hfill\break
R.Gregory, \prd 54 4955 1996.

\bibitem{GM} M.Barriola and A.Vilenkin, \prl 63 341 1989. \hfill\break
D.Harari and C.Lousto, \prd 42 2626 1990.

\bibitem{TUROK1} N.Turok, \prl 63 2625 1989.

\bibitem{TUROK2} N.Turok and D.Spergel, \prl 64 2736 1990.

\bibitem{NOTZ} D.N\"{o}tzold, \prd 43 R961 1991.

\bibitem{DUR} R.Durrer, M.Heusler, P.Jetzer and N.Straumann, \plb 259 48 1991. \hfill \break
R.Durrer, M.Heusler, P.Jetzer and N.Straumann, \npb 368 527 1992.

\bibitem{BV} M.Barriola and T.Vachaspati, \prd 43 1056 1991.

\bibitem{MICRO} D.P.Bennett and S.H.Rhie, \apj 406 L7 1993. \hfill \break
J.Borrill, E.J.Copeland, A.R.Liddle, A.Stebbins and S.Veeraraghavan, \prd 50 2469 1994. \hfill \break
R. Durrer, M.Kunz and A.Melchiorri, [astro-ph/9901377].

\bibitem{LESG} J.Scherk and J.Schwarz, \npb 81 223 1974. \plb 52 347 1974.
\hfill\break 
E.Fradkin and A.Tseytlin, \plb 158 316 1985. \hfill\break
C.Callan, D.Friedan, E.Martinec and M.Perry, \npb 262 593 1985. \hfill\break
C.Lovelace, \npb 273 413 1985.

\bibitem{JBD} P.Jordan, \zphys 157 112 1959.\hfill\break
C.Brans and R.H.Dicke, \pr 124 925 1961.

\bibitem{LSTRD} C.Gundlach \& M.Ortiz, \prd 42 2521 1990. \hfill\break
O.Pimental \& A.N.Morales, {\it Rev.\ Mex.\ Fis.\ }{\bf 36} S199 (1990).
\hfill\break 
M.E.X.Guimaraes, \cqg 14 435 1997. \hfill \break
R.Gregory and C.Santos, \prd 56 1194 1997.

\bibitem{GSTRD} A.A.Sen, N.Banerjee and A.Banerjee, \prd 56 3706 1997.

\bibitem{GSTRDUS} O.Dando and R.Gregory, \prd 58 023502 1998.

\bibitem{GMD} A.Barros and C.Romero, \prd 56 6688 1997. 
\hfill\break
A.Banerjee, A.Beesham, S.Chatterjee, A.A.Sen, \cqg 15 645 1998.

\bibitem{GMDUS} O.Dando and R.Gregory, \cqg 15 985 1998.

\bibitem{PENSEL} U.Pen, U.Seljak and N.Turok, \prl 79 1611 1997.

\bibitem{DAMVIL} T.Damour and A.Vilenkin, \prl 78 2288 1997. 

\end{thebibliography}
\end{document}